# The Naked Truth about Hadronic Regge Trajectories

A.E. Inopin

Mathem. Department, Tver State University, Sadovyj per.,35, 170002, Tver, Russia,
& Virtual Teacher Ltd., Vancouver, Canada.

We have reconstructed Regge trajectories (RT), using all existing data on masses and spin-parities of all currently known hadrons. In this resonance energy region meson and baryon RT are grossly nonlinear, and only 12% of all RT could be classified as a linear, with $\sigma \sim 0$, $\alpha' \approx 0.9$ GeV$^{-2}$.

## 1. INTRODUCTION

Regge trajectories (RT) in hadron physics have been known for some 40 years. Initially they were introduced by Tullio Regge [1,2], who simply generalized the solution for scattering amplitude by treating the angular momentum $L$ as a complex variable. He proved that for a wide class of potentials the only singularities of the scattering amplitude in the complex $L$ plane were poles, now called Regge poles. If these poles occur for positive integer values of $L$ they correspond to the bound states or the resonances, and they are also important for determining certain technical aspects of the dispersion properties of the amplitudes. Regge interpreted the simple poles of $a_l(k^2)$ on the complex $L$-plane to be either resonances or bound states. Chew and Frautschi [3] applied the Regge poles theory to investigate the analyticity of $a_l(k^2)$ in the case of strong interactions. They simply postulated that all strongly interacting particles are self-generating (the bootstrap hypothesis) and that they must lie on Regge trajectories (Chew-Frautschi conjecture) [4]. At first, linearity was just a convenient guide in constructing the Chew-Frautschi plots, because data were scarce and there were few *a priori* rules to direct the mesons and baryons into the same trajectories [5]. Once linearity was found to be a good working hypothesis, justification was given through certain assumptions in the Regge poles theory as follows: For Re$L \geq -1/2$, the partial-wave components of the scattering amplitude f have only simple poles and are functions of $k^2$

$$a_l(k^2) \approx \beta(k^2) / (L - \alpha(k^2)), \qquad (1)$$

where $\beta$ is the residue and $\alpha$ the position (Regge trajectory) of the simple poles.

By the end of the 1960s quarks were discovered experimentally and quark-parton model emerged almost immediately [6]. In the 1970s quantum chromodynamics (QCD) got a firm ground as a theory of strong interactions. This theory, QCD, has nothing to do with the original framework of Regge and Chew-Frautschi, since it's dealing with different dynamical equations.

The aim of the present paper is to dissect the naked truth about hadronic Regge trajectories. The whole issue is an eclectic mix of confusions, partly because of a huge number of quark models, which are in many ways contradictory to each other. This leads to am-biguous conclusions about the true nature of Regge trajectories. In this situation the hadronic data itself presents the purest imprint of the hadronic world. Therefore we will scrutinize the last issue of Review of Particle Physics 2000 [7], and reconstruct all possible RT for mesons and baryons. Then we will extract all the slopes, characterizing the given RT and examine how they deviate from the standard recipe $\alpha' = 0.9$ GeV$^{-2}$. The best parameters which describe wild deviations of RT from linear and parallel lines are the dispersion $\sigma$ and the average value of slope for the given RT, $<\alpha'>$.

## 2. MESONS

We will work with the full listings of PDG [7]. First on the list is the light unflavored mesonic sector, where massive experimental discoveries were made during the last decade. The most extensively investigated are scalar-isoscalar f mesons, total 28 states. From these data [7,8] we could construct four radial and two orbital f trajectories. Radial RT for $f_2$ ($J^{pc} = 2^{++}$) is a 13-plet and it is the most nonlinear hadronic RT, with $\sigma =7.91$ GeV$^{-2}$ and $<\alpha'> = 6.37$ GeV$^{-2}$ (see Fig.1). This trajectory has two peak slopes of 17.54 GeV$^{-2}$ and 27.47 GeV$^{-2}$. The $f_0$ radial RT is nonet and it is also quite nonlinear , with $\sigma =1.69$ GeV$^{-2}$, $<\alpha'> = 2.30$ GeV$^{-2}$ with peak slope value of 6.10 GeV$^{-2}$. The $f_4$ radial RT has only three states, but it's essentially nonlinear, with $\sigma =2.85$ GeV$^{-2}$, $<\alpha'> = 4.40$ GeV$^{-2}$ and peak slope value of 6.41 GeV$^{-2}$. All the essentially nonlinear mesonic RT will be assembled in Table1. With new PWA just coming from Crystal Barrel data [8], it is possible to construct radial RT for the $f_1$ mesons ($J^{pc} = 1^{++}$). This $f_1$ is a quartet including the newly discovered $f_1$(1971). It is essentially nonlinear with $\sigma =1.55$ GeV$^{-2}$, $<\alpha'> = 2.29$ GeV$^{-2}$ with peak slope value of 3.69 GeV$^{-2}$. Orbital RT for f mesons include parent $f_0$ and daughter $f_0$. Parent $f_0$ is a quartet with $\sigma = 0.65$ GeV$^{-2}$, $\langle\alpha'\rangle = 1.29$ GeV$^{-2}$ and with peak slope value of 2.03 GeV$^{-2}$ and it is essentially nonlinear. Daughter $f_0$ RT is a triplet with $\sigma = 0.82$ GeV$^{-2}$, $\langle\alpha'\rangle = 1.26$ GeV$^{-2}$ and peak slope value of 1.84 GeV$^{-2}$. These two trajectories are essentially nonlinear and nonparallel.

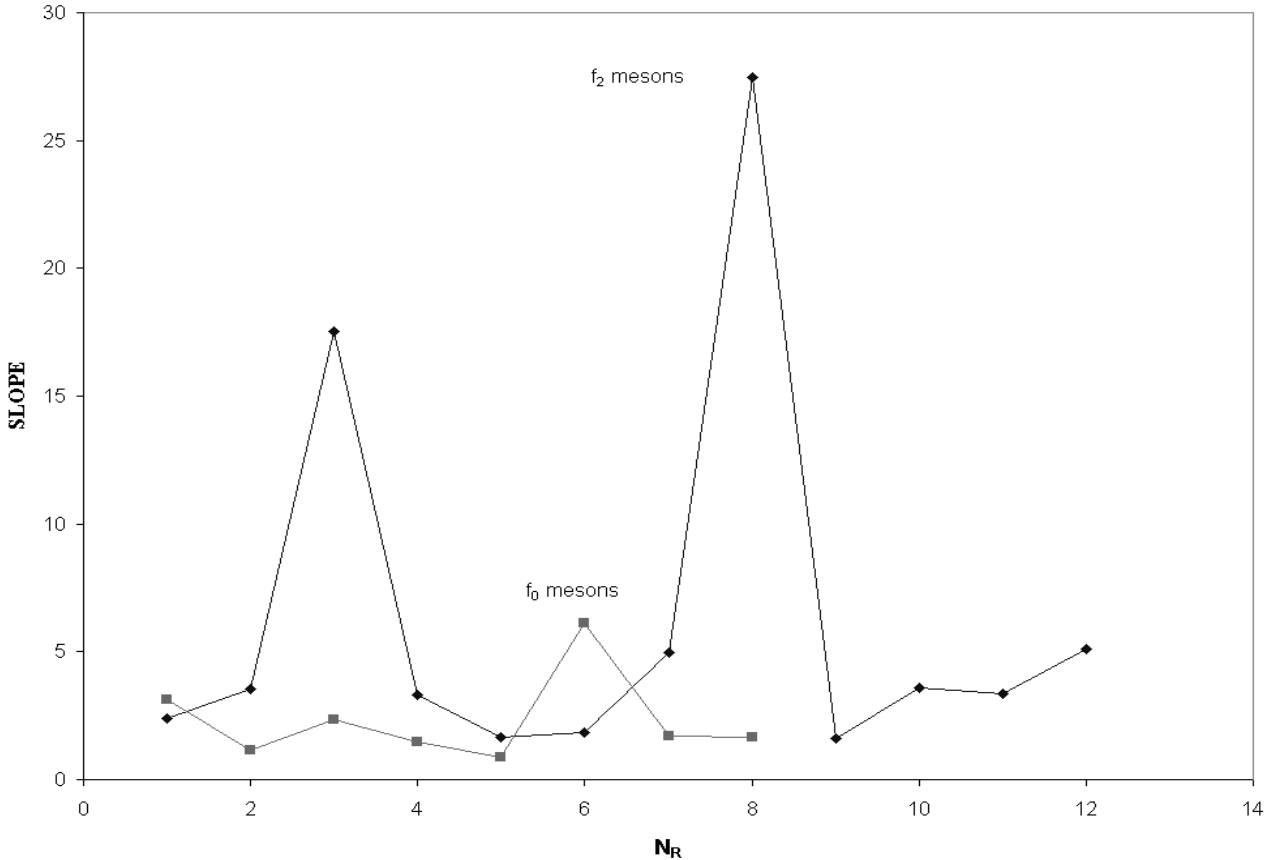

Fig.1. Slopes versus radial quantum number $N_r$ for $f_0$ and $f_2$ radial RT.

We now turn to the a-mesons. From the data [7,8] we could construct 4 radial and 3 orbital a-trajectories. The $a_0$ radial RT is a triplet with $\sigma = 0.20$ GeV$^{-2}$, $\langle\alpha'\rangle = 0.69$ GeV$^{-2}$. This trajectory is only slightly nonlinear. The $a_1$ radial RT is a quartet with $\sigma = 0.19$ GeV$^{-2}$, $\langle\alpha'\rangle = 0.79$ GeV$^{-2}$ and it's only slightly nonlinear. The $a_2$ radial RT is a sextet with $\sigma = 0.90$ GeV$^{-2}$, $\langle\alpha'\rangle = 1.73$ GeV$^{-2}$ with peak slope value of 3.18 GeV$^{-2}$. This trajectory is essentially nonlinear. The $a_3$ radial RT is a doublet with $\alpha' = 1.73$ GeV$^{-2}$.

Orbital a-trajectories are unique, because $a_0$ has parent, daughter and granddaughter RT. The $a_0$ parent RT is a quartet with $\sigma = 0.96$ GeV$^{-2}$, $\langle\alpha'\rangle = 1.50$ GeV$^{-2}$, with peak slope value of 2.61 GeV$^{-2}$. The $a_0$ daughter RT is a triplet with $\sigma = 1.05$ GeV$^{-2}$, $\langle\alpha'\rangle = 2.32$ GeV$^{-2}$, with peak slope value of 3.06 GeV$^{-2}$. The $a_0$ granddaughter RT is a triplet with $\sigma = 1.43$ GeV$^{-2}$, $\langle\alpha'\rangle = 3.88$ GeV$^{-2}$, with peak slope value of 4.88 GeV$^{-2}$. It's interesting that slopes are increasing successively from parent to daughter to granddaughter $a_0$ RT, with all three RT being nonparallel and essentially nonlinear.

Next on our list will be $h_1$ mesons. Combining data [7,8] with the just discovered at BNL $h_1(1594)$ [9], we have a quintet of states lying on one radial RT. This trajectory has $\sigma = 0.55$ GeV$^{-2}$, $\langle\alpha'\rangle = 1.25$ GeV$^{-2}$ with peak slope value of 1.81 GeV$^{-2}$, which makes it essentially nonlinear.

Table1: Slopes for essentially nonlinear meson RT ($\alpha'$, average $\langle\alpha'\rangle$, mean square deviation $\sigma$, in GeV$^{-2}$)

| RT for mesons | Slopes $\alpha'$ for neighbor pairs | $\langle\alpha'\rangle$ | $\sigma$ |
|---|---|---|---|
| $f_0(0^{++})$ parent | 3.00 0.78 0.94 | 1.58 | 1.01 |
| $f_0(0^{++})$ daughter | 1.84 0.68 | 1.26 | 0.82 |
| $f_0(0^{++})$ radial | 3.13 1.16 2.34 1.45 0.88 6.10 1.72 1.66 | 2.30 | 1.69 |
| $f_1(1^{++})$ radial | 2.56 3.69 0.63 | 2.29 | 1.55 |
| $f_2(2^{++})$ radial | 2.39 3.56 17.54 3.33 1.64 1.82 4.95 27.47 1.61 3.60 3.36 5.13 | 6.37 | 7.91 |
| $f_4(4^{++})$ radial | 2.38 6.41 | 4.40 | 2.85 |
| $a_2(2^{++})$ radial | 0.98 3.18 1.18 2.04 1.27 | 1.73 | 0.90 |
| $a_0(0^{++})$ parent | 2.61 0.86 1.03 | 1.50 | 0.96 |
| $a_0(0^{++})$ daughter | 3.06 1.58 | 2.32 | 1.05 |
| $a_0(0^{++})$ gr.daugh | 4.88 2.87 | 3.88 | 1.43 |
| $\eta$ radial | 0.72 2.65 0.96 1.06 1.10 3.70 | 1.70 | 1.20 |
| $h_1$ radial | 1.81 1.61 0.69 0.87 | 1.25 | 0.55 |
| $K(0^-)$ parent | 0.73 1.14 0.34 1.25 | 0.87 | 0.42 |
| $K(0^-)$ daughter | 1.69 2.38 | 2.04 | 0.49 |
| $K(0^-)$ radial | 0.53 1.92 1.44 | 1.30 | 0.71 |
| $K(1^+)$ radial | 2.90 1.32 | 2.11 | 1.12 |
| $K(2^-)$ radial | 1.54 6.49 0.57 | 2.87 | 3.18 |
| $J/\psi$ radial | 0.25 1.60 0.47 1.03 0.46 | 0.76 | 0.55 |
| $\chi_c(1P)$ parent | 1.51 3.11 | 2.31 | 1.13 |
| $\chi_b(1P)$ parent | 1.54 2.54 | 2.04 | 0.71 |
| $\chi_b(2P)$ parent | 2.11 3.66 | 2.89 | 1.10 |
| $\Upsilon$ radial | 0.09 0.15 0.21 0.16 0.30 | 0.18 | 0.08 |

There is no big news in pion sector. We have three radial RT for $\pi$ mesons. The $\pi$ radial RT ($J^{pc} = 0^{-+}$) is a triplet with $\sigma = 0.03$ GeV$^{-2}$, $<\alpha'> = 0.62$ GeV$^{-2}$ and it's practically linear. The $\pi_1$ radial RT is a doublet with $\alpha' = 1.55$ GeV$^{-2}$, which is quite large. The $\pi_2$ radial RT is a doublet with $\alpha' = 0.64$ GeV$^{-2}$. A plot of $\alpha'$ versus $M^2$ shows that all three RT are not parallel.

Let us consider $\eta$ mesons. With newly discovered states [8] we can construct two radial RT and one orbital RT. The $\eta$ radial RT is a septet with $\sigma = 1.20$ GeV$^{-2}$, $<\alpha'> = 1.70$ GeV$^{-2}$, with peak slope value of 3.70 GeV$^{-2}$. It is clearly essentially nonlinear RT. The $\eta_2$ radial RT is a quartet with $\sigma = 0.31$ GeV$^{-2}$, $<\alpha'> = 1.20$ GeV$^{-2}$, with peak slope value of 1.46 GeV$^{-2}$. This is fairly nonlinear RT, nonparallel to $\eta$. The $\eta$ parent orbital RT is a triplet with $\sigma = 0.08$ GeV$^{-2}$, $<\alpha'> = 0.79$ GeV$^{-2}$, and so it belongs to the class of linear trajectories.

We now turn to $\rho$ mesons. It is possible to construct one radial and one orbital RT. Radial RT is a quartet with $\sigma = 0.30$ GeV$^{-2}$, $<\alpha'> = 0.81$ GeV$^{-2}$, with peak slope value of 1.16 GeV$^{-2}$. This is fairly nonlinear RT. Parent orbital RT is a triplet with $\sigma = 0.07$ GeV$^{-2}$, $<\alpha'> = 0.83$ GeV$^{-2}$. This is almost linear RT.

Next on our list are $\omega$ mesons. Using data [7,8] we can construct only one radial RT for $\omega$ ($J^{pc} = 1^{--}$). This is a quintet with $\sigma = 0.29$ GeV$^{-2}$, $<\alpha'> = 1.07$ GeV$^{-2}$, with peak slope value of 1.42 GeV$^{-2}$. So it is slightly nonlinear radial RT.

The $\phi$-meson sector has quite sparse data with only one radial RT. It is a doublet with $\alpha' = 0.56$ GeV$^{-2}$.

Now we have finished an analysis of light unflavored mesons and turn to strange mesons ($S = \pm 1$, $C=B=0$). The K-mesons sector is very rich, encompassing 24 states so far. Even without appealing to exchange-degeneracy (EXD) we could construct a large number of trajectories. We have four radial and two orbital RT for K-mesons. $K(0^-)$ radial RT is a quartet with $\sigma = 0.71$ GeV$^{-2}$, $<\alpha'> = 1.30$ GeV$^{-2}$, with peak slope value of 1.92 GeV$^{-2}$. This is essentially nonlinear RT. $K(1^-)$ radial RT is a triplet with $\sigma = 0.16$ GeV$^{-2}$, $<\alpha'> = 0.94$ GeV$^{-2}$, with peak slope value of 1.06 GeV$^{-2}$. This is slightly nonlinear RT. $K(2^-)$ radial RT is a quartet with $\sigma = 3.18$ GeV$^{-2}$, $<\alpha'> = 2.87$ GeV$^{-2}$, with peak slope value of 6.49 GeV$^{-2}$. This is one of the most nonlinear RT in nature. $K(1^+)$ radial RT is a triplet with $\sigma = 1.12$ GeV$^{-2}$, $<\alpha'> = 2.11$ GeV$^{-2}$, with peak slope value of 2.90 GeV$^{-2}$. This is essentially nonlinear RT. All radial RT for K mesons, except $K(1^-)$ are essentially nonlinear.

Orbital K-meson's RT started from $K(0^-)$ parent spin singlet. It is a quintet with $\sigma = 0.42$ GeV$^{-2}$, $<\alpha'> = 0.87$ GeV$^{-2}$, with peak slope value of 1.25 GeV$^{-2}$. This is clearly essentially nonlinear RT. $K(0^-)$ daughter RT is a triplet with $\sigma = 0.49$ GeV$^{-2}$, $<\alpha'> = 2.04$ GeV$^{-2}$, which is essentially nonlinear. As we see, parent and daughter $K(0^-)$ trajectories are nonlinear and nonparallel. Vector $K^*(1^-)$ parent RT is a quintet with $\sigma = 0.13$ GeV$^{-2}$, $<\alpha'> = 0.84$ GeV$^{-2}$, and it is just slightly nonlinear RT.

Now we turn to charmed mesons ($C = \pm 1$) sector. There is insufficient data to construct RT there. The same situation persist in charmed-strange sector ($C=S=\pm 1$). Things are much more interesting in the charmonium sector. We have here one radial RT and one orbital RT. The radial trajectory, which started from famous $J/\Psi$, is a sextet with $\sigma = 0.55$ GeV$^{-2}$, $<\alpha'> = 0.76$ GeV$^{-2}$, with peak slope value of 1.60 GeV$^{-2}$. This is essentially nonlinear RT. The orbital trajectory is a parent $\chi_c(1P)$ and it is a triplet. It has $\sigma = 1.13$ GeV$^{-2}$, $<\alpha'> = 2.31$ GeV$^{-2}$, with peak slope value of 3.11 GeV$^{-2}$. This is essentially nonlinear RT. As we see, all the charmonium trajectories are essentially nonlinear.

The data on bottom mesons (B=±1), bottom, strange mesons (B=±1, S=±1) and bottom, charmed mesons (B=C=±1) are still insufficient to construct RT.

Next we consider bottomonium sector. From the data [7] we can construct one radial and two orbital RT. The $\Upsilon$ radial RT is a sextet with $\sigma$ =0.08 GeV$^{-2}$, $<\alpha'>$ = 0.18 GeV$^{-2}$. Although dispersion is small, this is strict consequence of small slopes, which are very different. This is essentially nonlinear RT. $\chi_b(1P)$ orbital RT is a triplet with $\sigma$ = 0.71 GeV$^{-2}$, $<\alpha'>$ = 2.04 GeV$^{-2}$. This is essentially nonlinear RT. The $\chi_b(2P)$ orbital RT is a triplet with $\sigma$ = 1.10 GeV$^{-2}$, $<\alpha'>$ = 2.89 GeV$^{-2}$, with peak slope value of 3.66 GeV$^{-2}$. This is essentially nonlinear RT, which is nonparallel to $\chi_b(1P)$.

We conclude that out of a total of 32 mesonic RT, 22 belong to the category of essentially nonlinear. Seven RT are fairly nonlinear, and only three RT are linear, which amounts to 9% share. (We did not account for doublets RT, which don't have a curvature).

## 3. BARYONS

### 3.1 N - $\Delta$

In the baryonic sector we have many more trajectories than for mesons. Our strategy will be to discuss the most interesting cases, leaving the rest for the tables and figures.

The nonstrange sector is very rich, comprising 23 N and 22 $\Delta$ states [7]. We will include in the analysis three new resonances, just discovered at ELSA, SAPHIR [7]: $D^*_{13}(1895)$, $S^*_{11}(1897)$ and $P^*_{11}(1986)$[1]. We will also include in analysis the so-called N(~3000 Region) and $\Delta$(~3000 Region), which are mostly the results of PWA by Hendry [11][2]. So, altogether we have 31 N and 28 $\Delta$ resonances. N, and $\Delta$ spectra exhibit a very interesting clustering structure. In nucleon sector we see the following four clusters: sextet $S_{11}(1650)$-$D_{15}(1675)$-$F_{15}(1680)$-$D_{13}(1700)$-$P_{11}(1710)$-$P_{13}(1720)$ is squeezed within 70 MeV interval; triplet $D^*_{13}(1895)$-$S^*_{11}(1897)$-$P_{13}(1900)$ is squeezed within 5 MeV interval; triplet $D_{13}(2080)$-$S_{11}(2090)$-$P_{11}(2100)$ is squeezed within 20 MeV interval, and quartet $G_{17}(2190$-$D_{15}(2200)$-$H_{19}(2220)$-$G_{19}(2250)$ is squeezed within 60 MeV interval. First cluster is split into three parity doublets: $S_{11}(1650)$-$P_{11}(1710)$, $D_{13}(1700)$-$P_{13}(1720)$, $D_{15}(1675)$-$F_{15}(1680)$. Second cluster has one parity doublet: $D^*_{13}(1895)$-$P_{13}(1900)$. Third cluster has one parity doublet: $S_{11}(2090$-$P_{11}(2100)$. Fourth cluster has one parity doublet: $H_{19}(2220)$-$G_{19}(2250)$.

In $\Delta$ sector we see the following two clusters: septet $S_{31}(1900)$-$F_{35}(1905)$-$P_{31}(1910)$-$P_{33}(1920)$-$D_{35}(1930)$-$D_{33}(1940)$-$F_{37}(1950)$ is squeezed within 50 MeV interval and triplet $F_{37}(2390)$-$G_{39}(2400)$-$H_{311}(2420)$ is squeezed within 30 MeV interval. First cluster is split into three parity doublets plus one extra state: $S_{31}(1900)$-$P_{31}(1910)$, $P_{33}(1920)$-$D_{33}(1940)$, $F_{35}(1905)$-$D_{35}(1930)$. Second cluster has no parity doublets. This clustering pattern and the precise mechanism of parity doubling in N, and $\Delta$ spectra remained the challenges for the current quark models. One promising approach introduced the so-called Rarita-Schwinger (RS) clusters [12]. As a result, the author found three RS clusters both in N and $\Delta$ spectra, as opposed to the experimentally seen four clusters in N, and two in $\Delta$ sectors.

Now we turn to the analysis of N, and $\Delta$ Regge trajectories. Major parent nucleon trajectory $P_{11}(938)$-$F_{15}(1680)$-$H_{19}(2220)$-$K_{113}(2700)$-$L_{117}(3500)$ is moderately nonlinear quintet, with $\sigma$ = 0.28 GeV$^{-2}$, $<\alpha'>$ = 0.81 GeV$^{-2}$. Major parent $\Delta$ trajectory $P_{33}(1232)$- $F_{37}(1950)$-$H_{311}(2300)$-$L_{315}(3700)$ is also just moderately nonlinear quartet, with $\sigma = 0.18$ GeV$^{-2}$, $\langle\alpha'\rangle = 0.78$ GeV$^{-2}$.

---

[1] We will mark by asterisk, "*", all the tentatively defined resonances throughout this paper.
[2] We include these data in our original papers [10], where we used NRQM to fit all the spectra.

Nevertheless there are plenty of nonlinear RT in the nonstrange sector (see Fig. 2). $P_{31}$ parent RT: $P_{31}(1750)$-$F_{35}(1905)$-$H_{39}(2300)$-$K_{313}(3200)$ is essentially nonlinear quartet with $\sigma = 1.63$ GeV$^{-2}$, $\langle\alpha'\rangle = 1.71$ GeV$^{-2}$, with peak slope value of 3.53 GeV$^{-2}$. $S_{31}$ parent RT: $S_{31}(1620)$-$D_{35}(1930)$-$G_{39}(2400)$-$I_{313}(2750)$-$L_{317}(3300)$-$N_{321}(4100)$ stretches to the highest possible mass 4.1 GeV. It is an essentially nonlinear sextet, with $\sigma = 0.56$ GeV$^{-2}$, $\langle\alpha'\rangle = 0.97$ GeV$^{-2}$, with peak slope value of 1.82 GeV$^{-2}$. $S_{11}$ parent RT: $S_{11}(1535)$-$D_{15}(1675)$-$G_{19}(2250)$ is extremely nonlinear triplet with $\sigma = 2.51$ GeV$^{-2}$, $\langle\alpha'\rangle = 2.67$ GeV$^{-2}$, and peak slope value of 4.44 GeV$^{-2}$. $D_{13}$ parent RT: $D_{13}(1520)$-$G_{17}(2190)$-$I_{111}(2600)$-$L_{115}(3100)$-$N_{119}(3750)$ is essentially nonlinear quintet, with $\sigma = 0.24$ GeV$^{-2}$, $\langle\alpha'\rangle = 0.74$ GeV$^{-2}$.

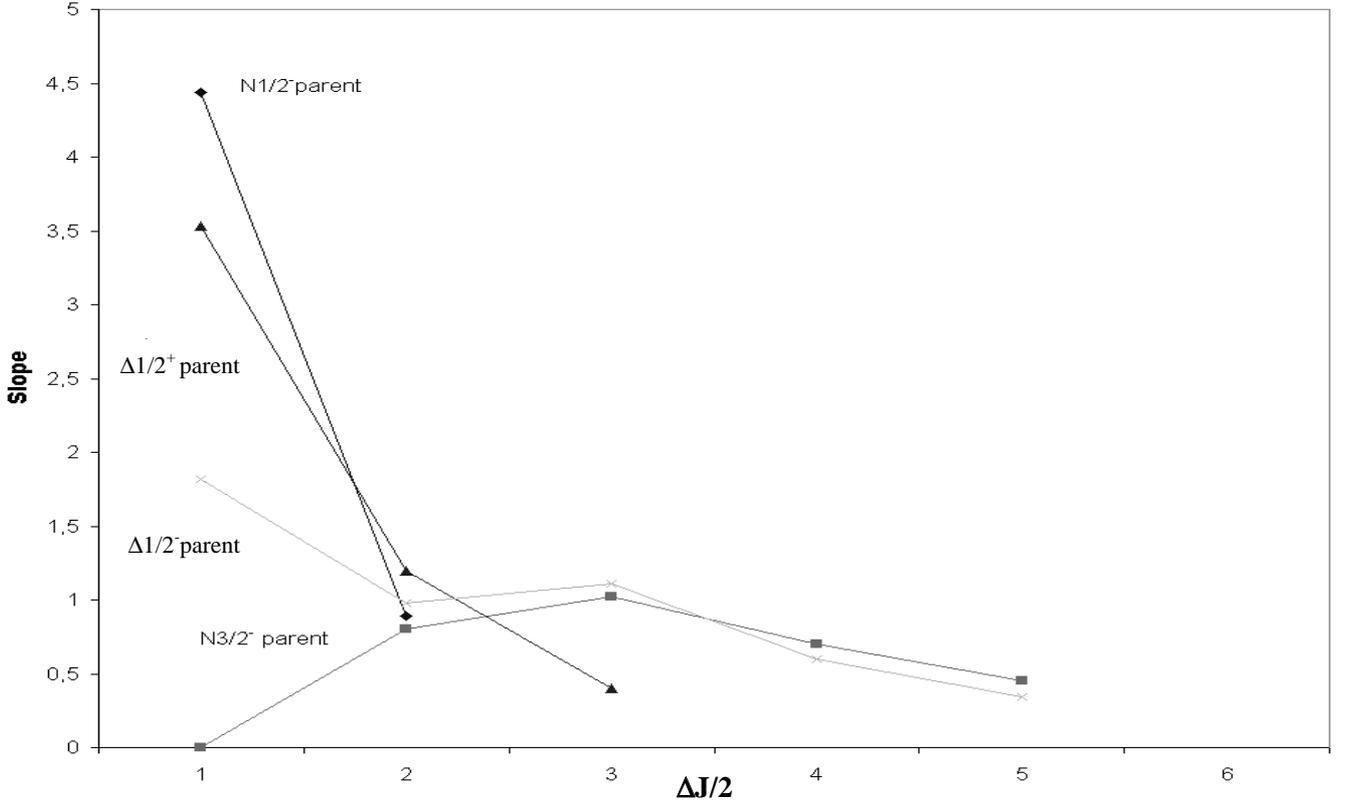

Fig. 2. Slopes for essentially nonlinear baryonic orbital RT.

Among the radial RT in N, $\Delta$ sector there are few essentially nonlinear. $S_{11}$ radial RT is a quartet: $S_{11}(1535)$-$S_{11}(1650)$-$S_{11}^*(1897)$-$S_{11}(2090)$. It is essentially nonlinear RT with $\sigma = 0.87$ GeV$^{-2}$, $\langle\alpha'\rangle = 1.72$ GeV$^{-2}$ and peak slope value of 2.72 GeV$^{-2}$. $D_{13}$ radial RT is a quintet: $D_{13}(1520)$-$D_{13}(1700)$-$D_{13}^*(1895)$-$D_{13}(2080)$-$D_{13}(2600)$. It is essentially nonlinear RT with $\sigma = 0.57$ GeV$^{-2}$, $\langle\alpha'\rangle = 1.23$ GeV$^{-2}$ and peak slope value of 1.72 GeV$^{-2}$. $P_{11}$ radial RT is a quintet: $P_{11}(939)$-$P_{11}(1440)$-$P_{11}(1710)$-$P_{11}^*(1986)$-$P_{11}(2100)$. It is essentially nonlinear RT with $\sigma = 0.59$ GeV$^{-2}$, $\langle\alpha'\rangle = 1.29$ GeV$^{-2}$ and peak slope value of 2.15 GeV$^{-2}$.

Some of the RT are too short (doublets) to judge on nonlinearity, but they have rather large slopes. $F_{35}$ radial RT is a doublet with $\alpha' = 2.7$ GeV$^{-2}$. $P_{31}$ daughter orbital RT is a doublet with $\alpha' = 5.68$ GeV$^{-2}$ and it is nonparallel to $P_{31}$ parent RT. $P_{13}$ parent orbital RT is a doublet with $\alpha' = 2.0$ GeV$^{-2}$. We conclude that in N, and $\Delta$ sector we have five nucleon and two delta essentially nonlinear RT.

Table 2: Slopes for essentially nonlinear baryon RT ($\alpha'$, average $\langle\alpha'\rangle$, mean square deviation $\sigma$, in GeV$^{-2}$)

| RT for baryons | Slopes $\alpha'$ for neighbor pairs | $\langle\alpha'\rangle$ | $\sigma$ |
|---|---|---|---|
| N$1/2^-$ parent | 4.44 0.89 | 2.67 | 2.51 |
| N$3/2^-$ parent | 0.80 1.02 0.70 0.45 | 0.74 | 0.24 |
| N$1/2^+$ radial | 0.84 1.18 0.98 2.15 | 1.29 | 0.59 |
| N$1/2^-$ radial | 2.72 1.14 1.30 | 1.72 | 0.87 |
| N$3/2^-$ radial | 1.72 1.43 1.36 0.41 | 1.23 | 0.57 |
| $\Delta 1/2^+$ parent | 3.53 1.20 0.40 | 1.71 | 1.63 |
| $\Delta 1/2^-$ parent | 1.82 0.98 1.11 0.60 0.34 | 0.97 | 0.56 |
| $\Lambda 1/2^+$ radial | 0.76 1.40 | 1.08 | 0.45 |
| $\Lambda 1/2^-$ radial | 1.23 2.22 1.32 | 1.59 | 0.55 |
| $\Lambda 3/2^-$ radial | 1.83 0.39 | 1.11 | 1.02 |
| $\Sigma 1/2^+$ radial | 0.83 7.58 2.65 2.49 | 3.39 | 2.91 |
| $\Sigma 1/2^-$ radial | 6.06 3.65 1.07 | 3.60 | 2.50 |
| $\Sigma 3/2^-$ radial | 3.41 1.03 | 2.22 | 1.68 |

### 3.2 $\Lambda$ - $\Sigma$

We turn now to the $\Lambda$-$\Sigma$ sector. There are many interesting features in this qqs sector. One of them is exchange degeneracy (EXD) hypothesis, which happened to hold quite well in the $\Lambda$-$\Sigma$ sector. As we will see later, EXD lead to trajectories with negative slopes, which never arise in N-$\Delta$ sector. Another feature is the clustering in the $\Lambda$-$\Sigma$ sector, which is qualitatively different from clustering in the N-$\Delta$ sector. Third feature is the existence of parity doublets in the $\Lambda$-$\Sigma$ sector. Because $\Lambda$-$\Sigma$ has only one strange quark, their shape is still not so deformed. If we can imagine that we have three balls in a bag, and two of them are of almost the same weight, while the third a bit heavier than the two, the bag will get the form of a pear. It will be reflectionally asymmetric. Near the rest the deformation is perhaps still not so dramatic and the lowest excitations are similar to those of the nonstrange baryons. For that reason we observe also in $\Lambda$-spectrum the same sequence $1/2^+$, $1/2^-$, $3/2^-$ as in nonstrange baryons. If the heavier ball starts to gain rotational energy, the deformation will increase. The pear shape gets more pronounced and when the pear oscillates it gives rise to parity doublets, which we already see.

Full listings [7] give to us 18 $\Lambda$ and 26 $\Sigma$ resonances. Some of the states are lacking the J$^P$ assignments. Let's take a closer look at this. State $\Lambda(2000)$ does not have J$^P$, but data from Cameron78 (see full listings [7]) allowed tentatively, the J$^P$ = $1/2^-$ assignment. Further evidence came from the recent paper by Iachello [13] and older one by Capstick-Isgur [14]. Therefore we assign J$^P$ = $1/2^-$ to the $\Lambda(2000)$. The $\Lambda$-states with highest masses, $\Lambda(2350)$ and $\Lambda(2585)$ were not described theoretically and there are no clear claims from the experiments. For this reason we will not include $\Lambda(2350)$, $\Lambda(2585)$ in our Regge analysis, and we have total of 16 $\Lambda$ resonances to work with.

The situation with $\Sigma$ hyperons is even more interesting. Two low-lying states, $\Sigma(1480)$ and $\Sigma(1560)$ do not have any J$^P$ assignments from the experiment and theory can't predict them either. We will exclude $\Sigma(1480)$, $\Sigma(1560)$ from our analysis. The production experiments [7] give strong evidence for $\Sigma(1620)$, tentatively claiming J$^P$ = $1/2^+$. This claim is in accord with fresh calculations by Iachello [13]. So with newly defined $\Sigma^*1/2^+(1620)$, we form an exact parity doublet $\Sigma 1/2^-(1620)$ - $\Sigma^*1/2^+(1620)$. The production experiments [7] give strong evidence for $\Sigma(1670)$ bumps without $J^P$ assignments. Using predictions by Iachello [13] and Isgur [14], we clearly get $J^P = 1/2^-$ for $\Sigma(1670)$. It's interesting that this way we have two resonances with the same mass and different $J^P$ (see [7]). Such a degeneracy waits proper theoretical explanation. The state $\Sigma(1690)$ has most likely claim from the data [7] as $J^P = 5/2^+$. We will assign $J^P = 5/2^+$ to $\Sigma(1690)$ in our analysis. Next $\Sigma$ hyperon without $J^P$ assignment will be $\Sigma(2250)$. Using the results from Iachello [13] and Isgur [14], we assigned $J^P = 5/2^-$ to $\Sigma(2250)$. Last few bumps, $\Sigma(2455)$, $\Sigma(2620)$, $\Sigma(3000)$ and $\Sigma(3170)$ has no experimental claims for $J^P$, and there are no theoretical predictions so far for such a high masses. For this reason we will not include $\Sigma(2455)$, $\Sigma(2620)$, $\Sigma(3000)$, $\Sigma(3170)$ in our analysis. Finally, we have total of 22 $\Sigma$ hyperons for our analysis.

Clustering pattern in $\Sigma$ spectrum is very nontrivial. We clearly see three clusters there. Quartet $P_{11}(1660)$-$D_{13}(1670)$-$S^*_{11}(1670)$-$F^*_{15}(1690)$ is squeezed within 30 MeV interval. There is one parity doublet within this cluster: $P_{11}(1660)$-$S^*_{11}(1670)$. Triplet $S_{11}(1750)$-$P_{11}(1770)$-$D_{15}(1775)$ is squeezed within 25 MeV interval. There is one parity doublet within this cluster: $S_{11}(1750)$-$P_{11}(1770)$. Triplet $F_{15}(2070)$-$P_{13}(2080)$-$G_{17}(2100)$ is squeezed within 30 MeV interval. There are no parity doublets in this cluster. It's amazing that we have three $1/2^+$ - $1/2^-$ parity doublets in the $\Sigma$ sector: $S_{11}(1620)$-$P^*_{11}(1620)$, $P_{11}(1660)$-$S^*_{11}(1670)$ and $S_{11}(1750)$-$P_{11}(1770)$.

In $\Lambda$ sector we witness only one cluster. This quartet $S_{01}(1800)$-$P_{01}(1810)$-$F_{05}(1820)$-$D_{05}(1830)$ is squeezed within 30 MeV interval. The whole cluster is split into two parity doublets: $S_{01}(1800)$-$P_{01}(1810)$ and $F_{05}(1820)$-$D_{05}(1830)$. Note that an author [12] suggested four clusters in $\Lambda$ sector. It is a big puzzle for current quark models to explain this difference in clustering and parity doubling between N-$\Delta$ and $\Lambda$-$\Sigma$ sectors.

Now we turn to Regge analysis of $\Lambda$-$\Sigma$ sector. We will mostly concentrate on essentially nonlinear trajectories. The $\Lambda 1/2^-$ radial RT is a quartet $S_{01}(1405)$-$S_{01}(1670)$-$S_{01}(1800)$-$S^*_{01}(2000)$. It is essentially nonlinear RT with $\sigma = 0.55$ GeV$^{-2}$, $<\alpha'> = 1.59$ GeV$^{-2}$ and peak slope value of 2.22 GeV$^{-2}$. The $\Lambda 3/2^-$ radial RT is a triplet $D_{03}(1520)$-$D_{03}(1690)$-$D_{03}(2325)$. It is essentially nonlinear RT with $\sigma = 1.02$ GeV$^{-2}$, $<\alpha'> = 1.11$ GeV$^{-2}$ and peak slope value of 1.83 GeV$^{-2}$. The $\Lambda 1/2^+$ radial RT is a triplet $P_{01}(1116)$-$P_{01}(1600)$-$P_{01}(1810)$. It is essentially nonlinear RT with $\sigma = 0.45$ GeV$^{-2}$, $<\alpha'> = 1.08$ GeV$^{-2}$. Many other $\Lambda$-trajectories possess some degree of nonlinearity. The $\Sigma 1/2^+$ radial RT is a quintet $P_{11}(1193)$-$P^*_{11}(1620)$-$P_{11}(1660)$-$P_{11}(1770)$-$P_{11}(1880)$. It is essentially nonlinear RT with $\sigma = 2.91$ GeV$^{-2}$, $<\alpha'> = 3.39$ GeV$^{-2}$ and peak slope value of 7.58 GeV$^{-2}$. The $\Sigma 1/2^-$ radial RT is a quartet $S_{11}(1620)$-$S^*_{11}(1670)$-$S_{11}(1750)$-$S_{11}(2000)$. It is essentially nonlinear RT with $\sigma = 2.91$ GeV$^{-2}$, $<\alpha'> = 3.39$ GeV$^{-2}$ and peak slope value of 6.06 GeV$^{-2}$. The $\Sigma 3/2^-$ radial RT is a triplet $D_{13}(1580)$-$D_{13}(1670)$-$D_{13}(1940)$. It is essentially nonlinear RT with $\sigma = 1.68$ GeV$^{-2}$, $<\alpha'> = 2.22$ GeV$^{-2}$ and peak slope value of 3.41 GeV$^{-2}$.

Amazingly, all essentially nonlinear radial RT in $\Lambda$-$\Sigma$ sectors are mirroring each other: $\Lambda 1/2^+$ - $\Sigma 1/2^+$; $\Lambda 1/2^-$ - $\Sigma 1/2^-$; $\Lambda 3/2^-$ - $\Sigma 3/2^-$. There are no essentially nonlinear RT among the orbital $\Lambda$-$\Sigma$ trajectories.

**3.3 EXD in $\Lambda$–$\Sigma$**

It has been known for years that exchange degeneracy seems to exist experimentally at least for strange hyperons [5]. We will construct appropriate RT with tentatively assigned resonances. Major $\Sigma$-trajectory in this scheme will be a quartet: $P_{11}(1193)$-$P_{13}(1840)$-$F^*_{15}(1690)$-$F_{17}(2030)$. This essentially nonlinear RT has one negative slope $\alpha_2' = -1.89$ GeV$^{-2}$, $\sigma = 1.47$ GeV$^{-2}$, $<\alpha'> = -0.20$ GeV$^{-2}$. Corresponding daughter RT is a triplet: $P^*_{11}(1620)$-$P_{13}(2080)$-$F_{15}(1915)$. This essentially nonlinear RT also has one negative slope $\alpha_2' = -1.52$ GeV$^{-2}$, $\sigma = 1.49$ GeV$^{-2}$, $<\alpha'> = -0.47$ GeV$^{-2}$, and it's nonparallel to the parent RT. Major negative parity $\Sigma$-trajectory is a quartet: $S_{11}(1620)$-$D_{13}(1580)$-$D_{15}(1775)$-$G_{17}(2100)$. This essentially nonlinear RT has one negative slope $\alpha_1' = -7.81$ GeV$^{-2}$, $\sigma = 5.32$ GeV$^{-2}$, $<\alpha'> = -1.67$ GeV$^{-2}$. Corresponding $\Sigma 1/2^-$ daughter RT is a triplet: $S^*_{11}(1670)$-$D_{13}(1940)$-$D^*_{15}(2250)$. This is moderately nonlinear RT with $\sigma = 0.18$ GeV$^{-2}$, $<\alpha'> = 0.90$ GeV$^{-2}$ and it's nonparallel to the parent RT.

We start $\Lambda$ sector with major $\Lambda 1/2^+$ trajectory, which is a quintet: $P_{01}(1116)$-$P_{03}(1890)$-$F_{05}(1820)$-$F_{07}(2020)$-$H_{09}(2350)$. This essentially nonlinear RT has one negative slope $\alpha_2' = -3.85$ GeV$^{-2}$, $\sigma = 2.76$ GeV$^{-2}$, $<\alpha'> = -0.71$ GeV$^{-2}$. Major negative parity $\Lambda 1/2^-$ trajectory is a quartet: $S_{01}(1405)$-$D_{03}(1520)$-$D_{05}(1830)$-$G_{07}(2100)$. This is essentially nonlinear RT with $\sigma = 1.20$ GeV$^{-2}$, $<\alpha'> = 1.64$ GeV$^{-2}$ and peak slope value of 3.03 GeV$^{-2}$.

As we see, EXD in $\Lambda$-$\Sigma$ sector leads to a new class of trajectories, which are characterized by negative average slopes. Five out of six $\Lambda$-$\Sigma$ trajectories are essentially nonlinear.

Table3: Slopes for baryonic essentially nonlinear EXD RT

| RT for baryons | Slopes $\alpha'$ for neighbor pairs | $<\alpha'>$ | $\sigma$ |
|---|---|---|---|
| $\Sigma 1/2^+$ parent | 0.51  -1.89  0.79 | -0.20 | 1.47 |
| $\Sigma 1/2^+$ daughter | 0.59  -1.52 | -0.47 | 1.49 |
| $\Sigma 1/2^-$ parent | -7.81  1.53  1.26 | -1.67 | 5.32 |
| $\Lambda 1/2^+$ parent | 0.43  -3.85  1.30  0.69 | -0.71 | 2.76 |
| $\Lambda 1/2^-$ parent | 3.03  0.96  0.94 | 1.64 | 1.20 |
| $\Xi 1/2^+$ parent | 0.48  3.14 | 1.81 | 1.88 |

### 3.4 $\Xi$, $\Omega$, Charmed, Beauty Baryons

We still have to analyze double-strange hyperons, $\Xi$ (qss). Full listings [7] give to us 11 $\Xi$'s. Some of the states are lacking the $J^P$ assignments. State $\Xi(1620)$ is a bump, which does not have $J^P$ assignment from the experiment, and it also could not be predicted by theory so far. For this reason, we will exclude $\Xi(1620)$ from our analysis. The state $\Xi(1690)$ does not have $J^P$ assignment from the experiment, but Iachello [13] predict this state with $J^P = 1/2^+$. The state $\Xi(1950)$ does not have $J^P$ assignment from the experiment, but Iachello [13] predict this state with $J^P = 3/2^+$. The state $\Xi(2030)$ has tentative assignment $J^P = 5/2^?$ from the experiment [7], and we will consider it as a $\Xi 5/2^+(2030)$. Resonance $\Xi(2120)$ does not have $J^P$ assignment from the experiment, but Iachello [13] predicts this state with $J^P = 3/2^-$. Next bump, $\Xi(2250)$ does not have $J^P$ assignment from the experiment, but Iachello [13] predicts $J^P = 1/2^+$. The state $\Xi(2370)$ does not have $J^P$ assignment from the experiment, but Isgur [14] predict this state with $J^P = 7/2^-$. The last resonance, $\Xi(2500)$, does not have $J^P$ assignment from the experiment, and it also could not be predicted by theory so far. For this reason, we will exclude $\Xi(2500)$ from our analysis. Finally, we have nine $\Xi$ resonances to work with: $P_{11}(1315)$-$P_{13}(1530)$-$P^*_{11}(1690)$-$D_{13}(1820)$-$P^*_{13}(1950)$-$F^*_{15}(2030)$-$D^*_{13}(2120)$-$P^*_{11}(2250)$-$G^*_{17}(2370)$. There is no clustering in $\Xi$ sector. Major parent Ξ RT is a doublet $P_{11}(1315)$-$F^*_{15}(2030)$, with $\alpha' = 0.84$ GeV$^{-2}$, $\sigma = 0$. Another orbital RT is a parent Ξ3/2$^-$. It is a doublet $D_{13}(1820)$-$G^*_{17}(2370)$, with $\alpha' = 0.87$ GeV$^{-2}$, $\sigma = 0$. Ξ1/2$^+$ radial RT is a triplet $P_{11}(1315)$-$P^*_{11}(1690)$-$P^*_{11}(2250)$. It has $\sigma = 0.31$ GeV$^{-2}$, $<\alpha'> = 0.67$ GeV$^{-2}$ and it is fairly nonlinear RT. The Ξ3/2$^+$ radial RT is a doublet $P_{13}(1530)$-$P^*_{13}(1950)$, with $\alpha' = 0.68$ GeV$^{-2}$, $\sigma = 0$. The Ξ3/2$^-$ radial RT is a doublet $D_{13}(1820)$-$D^*_{13}(2120)$ with $\alpha' = 0.85$ GeV$^{-2}$, $\sigma = 0$. As we see, basically all RT's in Ξ sector are too short to make a conclusions about their linear/nonlinear nature.

If we assume EXD for Ξ hyperons, we can construct parent Ξ1/2$^+$ triplet RT: $P_{11}(1315)$-$P^*_{13}(1950)$-$F^*_{15}(2030)$. This RT happens to be essentially nonlinear with $\sigma = 1.88$ GeV$^{-2}$, $<\alpha'> = 1.81$ GeV$^{-2}$ and peak slope value of $3.14$ GeV$^{-2}$. We will compile all EXD-essentially nonlinear RT in Λ, Σ, Ξ sectors in Table 3.

If we consider Ω, charmed baryons and beauty baryons, there are not enough data to construct and analyze Regge trajectories.

We conclude that out of total 21 baryonic RT 13 are essentially nonlinear (62%). Five RT are fairly nonlinear (24%), and only 3 trajectories (14%) are in fact linear. (We don't account for doublet RT here, because they have no curvature).

### 4. CONCLUSIONS

We have constructed and scrutinized here *all* possible RT for the full listings PDG2000 [7] and even accounted for the newest data on mesons [8,9]. We want to stress that this approach leads to minimal bias in the interpretation of the results, unlike the results from any available quark models.

We have shown that in the mesonic sector out of total 32 RT, 22 trajectories are essentially nonlinear (or 69%), and seven trajectories are fairly nonlinear (or 22%). Only three trajectories could be classified as linear, which amounts to 9% share. Among essentially nonlinear meson RT 10 are orbital and 12 are radial.

In baryonic sector out of total 21 RT, 13 trajectories are essentially nonlinear (or 62%), and five trajectories are fairly nonlinear (or 24%). Only three trajectories could be classified as linear, which amounts to 14% share. Among essentially nonlinear baryon RT four are orbital and 12 are radial.

Appropriate dispersion, $\sigma$, for nonlinear mesonic and baryonic RT span the range 1~7.9 GeV$^{-2}$, and slopes span range 1~27.5 GeV$^{-2}$.

We have four doublet RT in mesonic sector and 31 doublet RT in baryon sector, which don't have a curvature, and so this massive sector is used only for the evaluation of parallelism between different RT.

So, our results strongly disagree with general opinion, that hadron RT are straight and parallel lines. As the data shows, in the currently available resonance energy region, mesonic and baryonic RT are grossly nonlinear, and only small part (~12%) of all RT could be classified as linear, with $\sigma$~0, $\alpha'$≈0.9 GeV$^{-2}$.

The existence of clusters in baryon spectra and their absence in mesonic sector is a big puzzle. In N-sector we have four clusters: sextet, quartet and two triplets. In Δ-sector we have only two clusters: septet and triplet. In Σ-sector we have three clusters: quartet and two triplets. In Λ-sector we have one cluster, a quartet. The N, Δ clusters have average spacing between the levels of 8.9 MeV, and Λ, Σ clusters have very similar average spacing of 8 MeV.

First nucleon cluster is split into three parity doublets: $1/2^- - 1/2^+$, $3/2^- - 3/2^+$, $5/2^- - 5/2^+$.

First $\Delta$ cluster is also split into three parity doublets: $1/2^- - 1/2^+$, $3/2^- - 3/2^+$, $5/2^- - 5/2^+$. So they are exactly mirroring each other. Within nucleon clusters, parity doublets (PD) $1/2^- - 1/2^+$ and $3/2^- - 3/2^+$ occurred twice.

Second nucleon cluster has one PD: $3/2^- - 3/2^+$. Second $\Delta$ cluster has no PD at all.

Third nucleon cluster has one PD, $1/2^- - 1/2^+$ and fourth nucleon cluster has one PD, $9/2^- - 9/2^+$.

Within $\Sigma$ clusters we find two $1/2^- - 1/2^+$ PD, which are exactly mirroring two nucleon $1/2^- - 1/2^+$ PD.

In total, in N, $\Delta$, $\Lambda$, and $\Sigma$ sectors we have six $1/2^- - 1/2^+$ PD, three $3/2^- - 3/2^+$ PD, three $5/2^- - 5/2^+$ and one $9/2^- - 9/2^+$ PD. The exact dynamical reasons for such clustering and parity doubling patterns in baryons remains a big puzzle for theory.

**Acknowledgements**

Author is very grateful to G.S. Sharov and S. King for the help with manuscript.